\documentclass[conference]{IEEEtran}
\usepackage{cite}

\ifCLASSINFOpdf
  \usepackage[pdftex]{graphicx}
  \DeclareGraphicsExtensions{.pdf,.jpeg,.png}
\else
  \usepackage[dvips]{graphicx}
\fi
\usepackage{array}
\usepackage{fixltx2e}
\usepackage{dblfloatfix}

\usepackage{url}

\usepackage[utf8]{inputenc}

\usepackage{csquotes}
\MakeOuterQuote{"}

\begin{document}
%
\title{CYCLONE\\Unified Deployment and Management \\ of Federated, Multi-Cloud Applications}

\author{
\IEEEauthorblockN{Mathias Slawik\IEEEauthorrefmark{1}, Begüm İlke Zilci\IEEEauthorrefmark{2}, Yuri Demchenko\IEEEauthorrefmark{3}, José Ignacio Aznar Baranda\IEEEauthorrefmark{4},\\
Robert Branchat\IEEEauthorrefmark{5}, Charles Loomis\IEEEauthorrefmark{6}, Oleg Lodygensky\IEEEauthorrefmark{7}, Christophe Blanchet\IEEEauthorrefmark{8}}
\IEEEauthorblockA{\IEEEauthorrefmark{1}\IEEEauthorrefmark{2}SNET, Technische Universität Berlin, Germany, \{mathias.slawik,ilke.zilci\}@tu-berlin.de \\ 
\IEEEauthorrefmark{3} System and Network Engineering, University of Amsterdam, The Netherlands, y.demchenko@uva.nl \\
\IEEEauthorrefmark{4}DANA, i2cat Internet Research Center, Barcelona, Spain, jose.aznar@i2cat.net\\
\IEEEauthorrefmark{5}\IEEEauthorrefmark{6}SixSq Sàrl, Geneva, Switzerland, Email: \{rob,cal\}@sixsq.com}
\IEEEauthorrefmark{7}Lab. de l'Accelerateur Lin., Universite Paris Sud 11, Orsay, France, oleg.lodygensky@lal.in2p3.fr\\
\IEEEauthorrefmark{8} Institute of Bioinformatics, Gif-sur-Yvette, France, christophe.blanchet@france-bioinformatique.fr\\
}

%


\maketitle

\begin{abstract}
Various Cloud layers have to work in concert in order to manage and deploy complex multi-cloud applications, executing sophisticated workflows for Cloud resource deployment, activation, adjustment, interaction, and monitoring. While there are ample solutions for managing individual Cloud aspects (e.g. network controllers, deployment tools, and application security software), there are no well-integrated suites for managing an entire multi cloud environment with multiple providers and deployment models.
This paper presents the CYCLONE architecture that integrates a number of existing solutions to create an open, unified, holistic Cloud management platform for multi-cloud applications, tailored to the needs of research organizations and SMEs. It discusses major challenges in providing a network and security infrastructure for the Intercloud and concludes with the demonstration how the architecture is implemented in a real life bioinformatics use case.
\end{abstract}


%
\IEEEpeerreviewmaketitle

\section{Introduction}

Contemporary Cloud Computing architectures are layered. Basically, a \textit{Cloud infrastructure} executes a set of \textit{virtual machines} which are connected by (nowadays virtualized) \textit{networks}. On these VMs,  \textit{computing platforms} provide abstractions and a unified management of \textit{cloud applications}, offering many benefits, e.g., scalability, on-demand provisioning and portability. Futher layered cloud models can be found within the NIST Cloud Computing Reference Architecture \cite{NISTCCRA}, as well as the Intercloud Architecture Framework (ICAF) \cite{DNL13}.

Clouds are managed in a highly automated fashion by a number of common tools, e.g., PaaS platforms, deployment managers, identity and access managers, as well as network controllers. Furthermore, many applications are federated over multiple Clouds, adding to the architecture complexity.

The "Intercloud" is trending: a globally integrated Cloud of Clouds sharing APIs, protocols, and data formats. A proposal to use existing standards and common mechanisms to achieve the "Intercloud Root" was made by Bernstein, et al. in \cite{BLSDM09}. Our previous work includes the definition of the ICAF in \cite{DNL13}, adressing Intercloud issues, as well as defining models and architecture patterns for federated access control within Intercloud environments in \cite{DNL14}.

A future Intercloud could incorporate tightly integrated Cloud platforms such as OpenStack\footnote{https://www.openstack.org} and the VMWare Software-Defined Data Center\footnote{http://www.vmware.com/software-defined-datacenter/}. While these tools provide integrated Cloud management, they fail to deliver open and standardized APIs, protocols, and data formats, and their components are difficult to replace. Furthermore, Intercloud scenarios require cross-cloud (e.g., public-private) interoperability, compatibility, and interchangeability which is currently challenging to implement. Thus, Application Service Providers (ASPs) as well as their customers are constricted in their deployment of well-integrated Cloud solutions, and the Intercloud awaits its implementation in practice.

The \textbf{CYCLONE} project (\textbf{C}omplete and D\textbf{y}namic Multi-\textbf{clo}ud Applicatio\textbf{n} Manag\textbf{e}ment) aims to enhance end-to-end cloud security and facilitates the deployment, management, and use of complex, multi-cloud applications. The project improves existing production-quality tools to offer a software stack based on open standards and APIs. Concrete project tasks are: Network-IaaS integration, common federated authentication, Intercloud brokering and matchmaking, as well as providing PaaS functionality to ASPs (e.g., VM coordination and automated placement). This publication presents the CYCLONE stakeholders and their requirements, a description of the CYCLONE architecture, as well as an evaluation in form of a practical deployment within the bioinformatics sector.

\section{Stakeholders and Requirements}
\label{ch:requirements}

There are four main \textbf{CYCLONE stakeholders}: \textit{(1) cloud application service providers} will use CYCLONE components to offer diverse functionality to \textit{(2) cloud developers} and \textit{(3) cloud operators} to enable them to easily implement, deploy, and manage cloud applications for \textit{(4) cloud application end-users}. Nine main requirements for CYCLONE components' functionality are described in the remainder of this chapter, which are pertinent to complex, federated, multi-cloud applications.

\textbf{Federated identity (R1):} Multi-cloud environment security requires a common authentication system where Cloud application end-users can log in to applications using a federated identity.

\textbf{End-to-end encryption (R2):} Common HTTP intermediaries, such as reverse proxies and load balancers, often act as TLS server connection ends, accessing HTTP/TLS plaintext. There has to be an end-to-end encryption (i.e., user agent to origin server) of sensitive HTTP entity bodies.

\textbf{Distributed logging (R3):} Managing multi-cloud applications effectively requires consolidating all event logs of the involved software, i.e., a distributed logging system.

\textbf{Deployment description (R4):} There has to be a method of describing Cloud application deployments, at best supporting scripting, multi-cloud deployment and orchestration, as well as custom application lifecycle hooks.

\textbf{Management APIs (R5)} Any IaaS solution should provide an easy to use, comprehensible, and rich set of management APIs (R5) for computing, storage, and network.

\textbf{VM marketplace (R6)} To allow collaboration between end-users, VM appliance creators and Cloud administrators, the IaaS solution should incorporate a comprehensible VM marketplace.

\textbf{Network service management platform (R7):} As cloud applications have limited network control and visibility, it is challenging to achieve service delivery automation, resource management, and on-demand network connectivity. To implement advanced network services, there has to be a network service management platform, integrated into the employed IaaS offering.

\textbf{Brokering (R8):} In order to support Cloud application developers in finding suitable services in the vast Intercloud, there has to be a service formalization, a service vocabulary, and a brokering component. It should adapt to dynamically changing Intercloud services' properties.

\textbf{Matchmaking (R9):} There should be automated matchmaking of Intercloud services properties to cloud operator requirements in order to guide service assessment and selection.

\section{Related Work}
\label{ch:related-work}

This chapter briefly presents the work related to CYCLONE, divided into areas pertinent to different CYCLONE components - in the order of the cloud stack, from infrastructure to application.

\textbf{IaaS platforms.} The most prominent IaaS ecosystem is Amazon EC2, a proprietary public Cloud platform. Two strong open source contenders are OpenStack and OpenNebula \cite{milojivcic2011opennebula}. However, they encompass a multitude of components whose set-up requires Cloud operators to follow extensive installation guides.

\textbf{Configuration Management.} The most commonly used configuration management tools are Chef\footnote{www.chef.io} and Puppet\footnote{puppetlabs.com}, both Ruby-based. They require the installation of a server and client agents. Juju\footnote{https://jujucharms.com/} is a notable tool for deploying and managing Cloud services, featuring a high level application model.

\textbf{Distributed logging.} The most widely deployed distributed logging system is \texttt{syslog}, standardized in \cite{rfc5424}. Syslog is often used for lower-level logging, e.g., logging of network daemon output. State-of-the art higher-level application logging includes Logging-as-a-Service offerings, such as Loggly\footnote{https://www.loggly.com/} and Papertrail\footnote{https://papertrailapp.com/}, as well as open source stacks, such as the ELK-stack\footnote{https://www.elastic.co/products}, which consists of Elasticsearch (persistence), Logstash (logging middleware) and Kibana (logging dashboard).

\textbf{Software-defined-Networking} Software Defined Networking (SDN) enables dedicated SDN controller software to take all network decisions and has recently gained acceptance\cite{SDN}. SDN eases data centre (DC) network management and enables new networking capabilities, e.g., centralized control, decoupling of network planes, exposing network APIs, and more. While SDN was initially proposed for core networks, IaaS providers are adopting it to allow holistic infrastructure management (IT+Network). The most relevant open source SDN solutions are OpenDayLight\footnote{http://www.opendaylight.org/} (ODL), RYU\footnote{http://osrg.github.io/ryu/}, and ONOS\footnote{http://onosproject.org/}. ODL is a popular solution supporting many network technologies. Some Cloud research projects, e.g., COSIGN\footnote{http://www.fp7-cosign.eu/} and BEACON\footnote{http://www.beacon-project.eu/}, are using it as intra-DC network management system. Yet, it is heavily influenced by vendors, had significant changes, lacks clear documentation, and has a very large codebase. RYU is an open source project initiated by NTT. It is easy to start and well documented. There is limited support of different network devices as RYU focuses on OpenFlow\footnote{https://www.opennetworking.org/sdn-resources/openflow/} (OF). Also, as OF does not support distributed DCs, Cloud federation scenarios are not supported. ONOS, created by ON.lab, is an open source SDN controller focusing on network performance, scalability, and multilayer designs. Despite being first released in December 2014, it seems to be a promising SDN platform.

All major vendors have commercial SDN solutions: Juniper OpenContrail, Cisco Application Centric Infrastructure, HP SDN VAN controller, Brocade SDN controller (formerly Vyatta), as well as the VMware NSX platform.

\textbf{End-to-end encryption.} There are different technologies for securing HTTP entity bodies end-to-end, notably Secure/Multipurpose Internet Mail Extensions (S/MIME) \cite{rfc5751}, XML Encryption \cite{XMLENC} and Signature \cite{XMLDSIG}, HTTPSec \cite{HTTPSEC}, as well as Secure HTTP (S-HTTP) \cite{rfc2660}. In one of our previous publications we analyze them, show the severe deficiencies of these technologies in Cloud environments and create a novel end-to-end HTTP-based encryption protocol, the Trusted Cloud Transfer Protocol (TCTP) \cite{S13d}.

\textbf{Federated Identity.} There are a number of existing federated identity patterns which can be considered state-of-the-art: One of the most comprehensible is the "Claims Based Identity \& Access Control"\footnote{http://claimsid.codeplex.com/} which uses SAML for transmitting security claims between participating parties. Another more recent approach to federated identity is OpenID Connect\footnote{http://openid.net/connect/} which defines a "simple identity layer" based on OAuth 2.0\footnote{http://oauth.net/2/} and JSON Web Token\footnote{http://jwt.io/}. OpenID Connect is notable for its wide industry adoption, e.g., by Google, Microsoft, and Salesforce, as well as its ongoing adoption by the GSMA to be used worldwide by mobile network operators.

\textbf{Complex Application Descriptions.} CYCLONE performed an analysis of contemporary application models, i.e., SlipStream, CIMI~\cite{cimi}, TOSCA~\cite{tosca}, OCCI~\cite{occi} and Compose YML~\cite{docker-yaml}. The model used by SlipStream already supports full multi-cloud application deployments and triggers. The DMTF CIMI model promises the best overall interoperability within the CYCLONE target area. Therefore, CYCLONE will extend it and use it for describing complex, federated, multi-cloud applications. The rationale can be found in \cite{deliverable_6_1}.

\begin{figure*}
\centering
\includegraphics[width=28cm, trim=0.65cm 28.3cm 3.7cm 2.45cm]{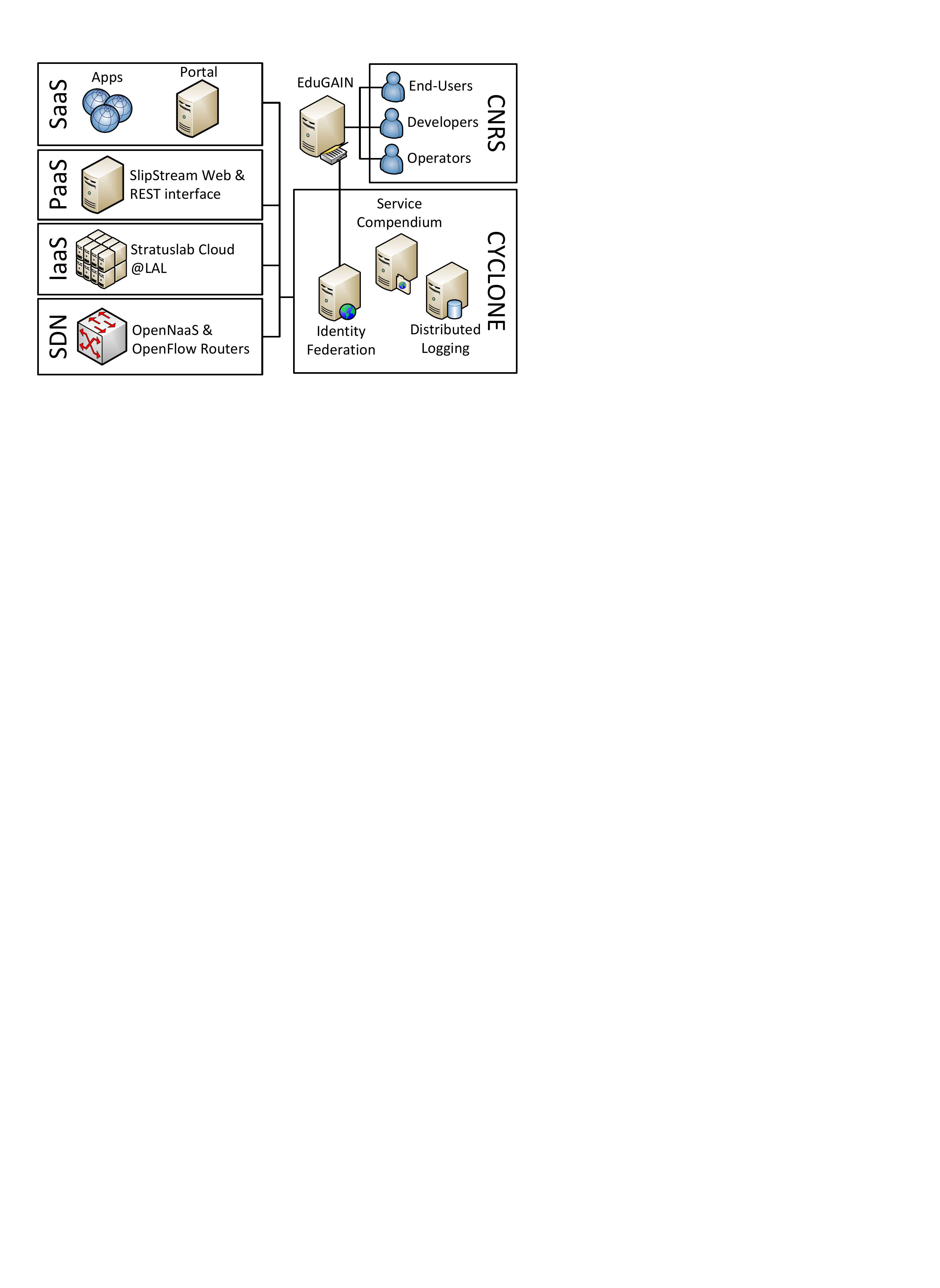}
\caption{The Bioinformatics Use Case}
\label{fig_sim}
\end{figure*}

\textbf{Service Matchmaking and Brokering.} There are interface description languages, such as WSDL \cite{WSDL}, as well as service description languages from the Semantic Web, e.g., OWL-S \cite{MPMB+05}, SAWSDL\cite{SAWSDL}, WSMO \cite{WSMO}, and Linked-USDL \cite{PCL14}. Most of the semantic approaches were abandoned, possibly due to missing industry adoption, e.g., OWL-S (2006), SAWSDL (2007), and WSMO (2008). In our previous work \cite{SK14} we have revealed unresolved challenges while building a Cloud service repository for regular Internet users: business pertinence, tooling simplicity, information reuse, and Cloud service modeling challenges. To adress these issues, we are building the Open Service Compendium (OSC), an open, wiki-like information system featuring a textual DSL, business-pertinent vocabulary, and a brokering component, presented in depth in \cite{SZKK15}.

In our survey \cite{zilci2015survey}, we observe that service matchmaking approaches from the academia have different problem definitions since they consider different service descriptions. Most of the related work addresses only numeric QoS parameters in a complex manner. However, important service selection criteria which are not numeric are not addressed e.g. list typed properties. In the CYCLONE architecture, we consider that service requesters specify the Clouds to use in advance, however the exact deployment actions are triggered in the background. Following the Grozev taxonomy \cite{grozev2014inter}, CYCLONE can be classified as a combination of trigger action and directly managed, while keeping the user's control of the Cloud distribution.

\section{CYCLONE Architecture and Components}
\label{ch:architecture}

\begin{figure}[h]
\centering
\includegraphics[width=\columnwidth, trim=0.35cm 28.3cm 17cm 2.45cm]{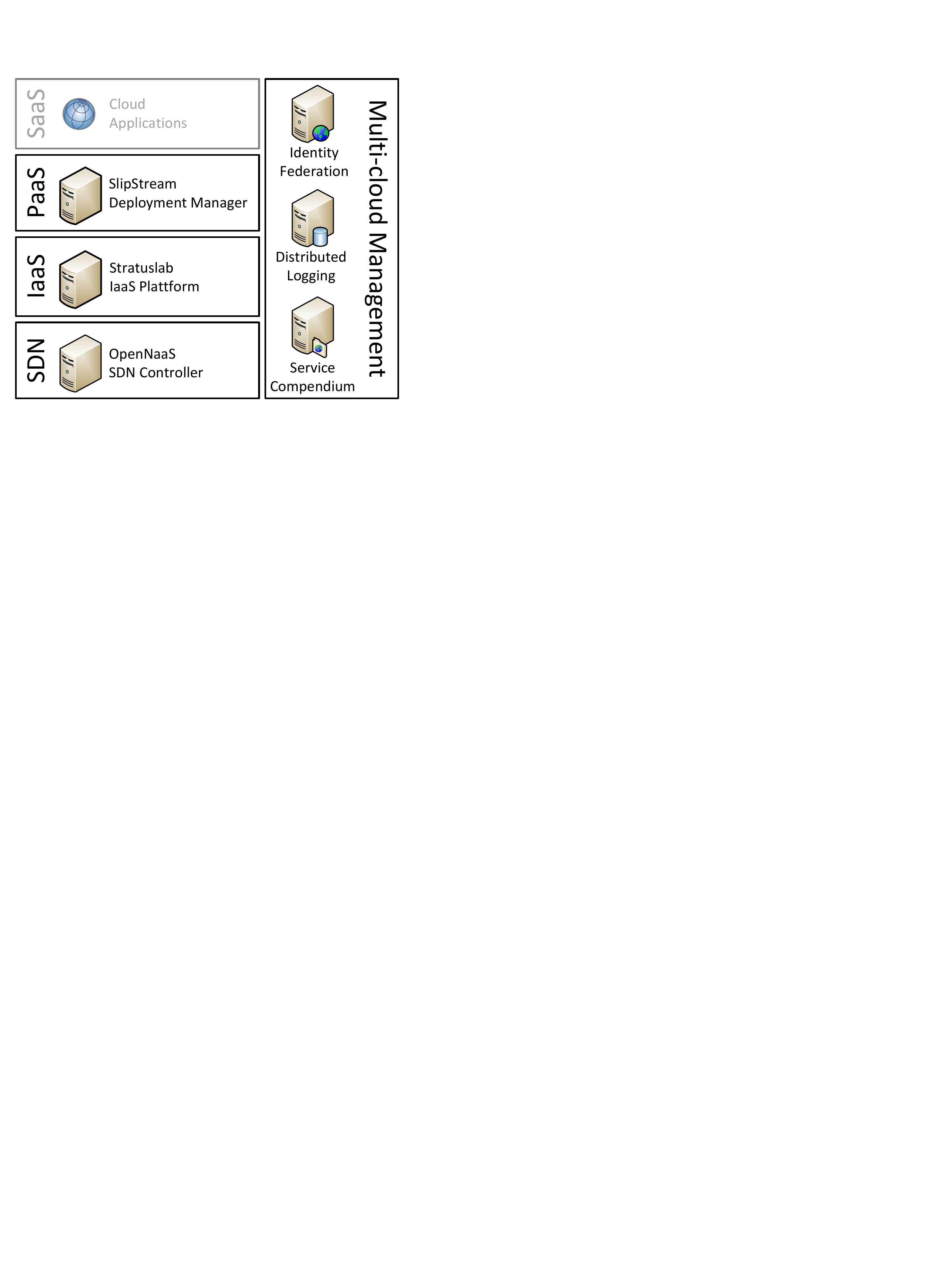}
\caption{The CYCLONE components}
\label{fig_components}
\end{figure}

The CYCLONE architecture comprises platform, infrastructure, networking, and multi-cloud management software and is presented in Figure \ref{fig_components}.

It is comprised of components for the \textbf{Cloud stack} (OpenNaaS, StratusLab, and SlipStream), as well as for overall \textbf{multi-cloud management} responsibilities (Identity Federation, Distributed Logging, and the Open Service Compendium).

\textbf{OpenNaaS} is a software-defined network controller for the IaaS interconnections. It provides scheduled, dynamic, and flexible network connectivity services with short provisioning times, thus reducing management complexity and bridging the Cloud to end-to-end network services, both intra- and inter-Cloud.

\textbf{StratusLab} provides infrastructure management APIs, comparable to OpenStack and Amazon EC2. It is easy to set up and features a substantially simplified architecture while being suitable for HPC workflows, as it is based on OpenNebula. Apart from compute, storage and network features, it can also be federated with other StratusLab installations. It uses OpenNaaS to manage the routers connecting the StratusLab VM hosts, offering network services to deployed VMs. As it is published as open source, it can be easily extended and integrated with other CYCLONE components.

\textbf{SlipStream} is an open source Cloud application manager.  It is used within CYCLONE to deploy Cloud applications onto one or more IaaS Cloud infrastructures and to manage the Cloud resources allocated to the running Cloud applications.  SlipStream combines its deployment engine with an "App Store", "Cloud Service Catalog", dashboard, and monitoring to provide a complete engineering PaaS supporting
DevOps processes. SlipStream supports most public IaaS services and IaaS Cloud distributions, including StratusLab.  In the interest of interoperability, SlipStream is gradually moving from its own RESTful API to the standardized CIMI API.

The \textbf{CYCLONE Identity Federation} is based on  Keycloak\footnote{http://keycloak.jboss.org/}, an open source SSO and IDM solution for browser apps and RESTful services. It is built on top of the OAuth 2.0, Open ID Connect, JSON Web Token (JWT) and SAML 2.0 specifications. CYCLONE end-users can use their own identity provider, e.g., EduGAIN, Shibboleth, Facebook, and Google, to log in to CYCLONE services. This also includes, for example, the login of Cloud developers into SlipStream and of Cloud operators into the distributed logging.

The \textbf{CYCLONE Distributed Logging} unifies the log messages of the Cloud stack and multi-cloud management components and offers end-users easy browsing through their logs. It is build upon the ELK logging stack.

The \textbf{Open Service Compendium} describes IaaS solutions which are available for application deployment. It presents Cloud operators the properties of different offerings, as well as a comparison of deployment options and their costs. We strive for SlipStream integration in order to raise the manageability of complex multi-cloud applications.

\section{Bioinformatics Use Case and evaluation}
\label{ch:use-case}
The first deployment of the CYCLONE architecture will be in a bioinformatics use case as shown in Figure \ref{fig_sim}, which is described in this section, along with the CYCLONE requirements which are achieved.

Cloud application users from a set of French laboratories use a \textbf{portal (R6)} to select specific Cloud applications for self-service deployment. This portal uses the SlipStream REST interface in order to deploy them on a StratusLab Cloud hosted by CNRS LAL. SlipStream utilizes a rich \textbf{deployment description (R4)} and consumes the StratusLab \textbf{VM management APIs (R5)}. Some of the applications are TCTP-enabled to achieve \textbf{end-to-end encryption (R2)}. The OpenNaaS \textbf{CYCLONE network management platform (R7)} configures Openflow-based routers used within the infrastructure at CNRS LAL. Applications and most components integrate with the CYCLONE federation provider allowing CNRS end-users to utilize their RENATER (EduGAIN) accounts for \textbf{federated identity (R1)}. All relevant components log their output to the CYCLONE \textbf{distributed logging (R3)}. Cloud Developers can use the Open Service Compendium for service \textbf{brokering (R8)} and \textbf{matchmaking (R9)}, possibly integrated with SlipStream.

\section{Conclusion}
\label{ch:conclusion}

This publication introduced the challenges, stakeholders and requirements of cloud application deployment and management in multi-cloud and multi-provider environments. To meet stakeholder requirements and associated challenges we introduce the CYCLONE architecture and demonstrate its benefits within the practical use case of bioinformatics applications. Through the integration and extension of specifically selected tools, we hope to create a simple, capable, and holistic Cloud management solution. In the future, we will further maturate our architecture and tooling to tackle open engineering challenges and further use cases. This should create an alternative to the complex and monolithic Cloud stacks prevalent today. Moreover, we plan to design test cases and document quantitative results to compare our approach to other Cloud stacks especially in the multi-cloud context. 


\section*{Acknowledgment}

This work is supported by the CYCLONE Horizon 2020 Innovation Action
CYCLONE \footnote{cyclone-project.eu}, funded by the European Commission under grant number 644925.

\IEEEtriggeratref{11}
\bibliographystyle{IEEEtran}
\bibliography{IEEEabrv,netcloud2015mathias,ilke,jose,cal}

\end{document}